\documentclass[12pt]{article}
\usepackage{amsfonts,amsbsy,latexsym,amssymb,amscd,amstext}
\newcommand{\be}{\begin{equation}}
\newcommand{\ee}{\end{equation}}
\newcommand{\bea}{\begin{eqnarray}}
\newcommand{\eea}{\end{eqnarray}}
\newcommand{\ben}{\begin{eqnarray}}
\newcommand{\een}{\end{eqnarray}}

\renewcommand{\>}{\rangle}

\newcommand{\noind}{\noindent}
\newcommand{\etac}{\!\!\ ^1S_0}
\newcommand{\jpsi}{\!\!\ ^3S_1}
\newcommand{\hypsp}{\!\!\ ^3S_1-\!\!\ ^1S_0}
\newcommand{\avs}{\overline S}
\newcommand{\avp}{\overline P}
\newcommand{\MeV}{{\rm MeV}}

\begin{document}

{\Large
\centerline{\bf
Quenched charmonium spectrum}}
\vskip 0.3cm

\noind QCD-TARO Collaboration:
        S.~Choe${}^a$,
        Ph.~de~Forcrand${}^{b,c}$,
        M.~Garc\'{\i}a~P\'erez${}^d$,
        Y.~Liu$^{\rm e}$,
        A.~Nakamura${}^f$,
        I.-O.~Stamatescu${}^{g,h}$,
        T.~Takaishi${}^i$
        and
        T.~Umeda${}^k$  
\vskip 0.4cm

\noind ${}^a$  Department of Chemistry, KAIST, 373-1 Kusung-dong, Yusung-gu,
Daejon 305-701, Korea\\
\noind ${}^b$ Institut f\"ur Theoretische Physik,
       ETH-H\"onggerberg, CH-8093 Z\"urich, Switzerland \\
\noind ${}^c$ Theory Division, CERN, CH-1211 Geneva 23, Switzerland \\
\noind ${}^d$ Instituto de F\'{\i}sica Te\'orica, Universidad Aut\'onoma de
Madrid, Cantoblanco, 28049 Madrid, Spain\\
\noind ${}^e$ Department of Physics, Nakai University, Tianjin 300071, China \\
\noind ${}^f$ IMC, Hiroshima University, Higashi-Hiroshima 739-8521, Japan\\
\noind ${}^g$ Institut f\"ur Theoretische Physik, Universit\"at
       Heidelberg,  D-69120 Heidelberg, Germany \\
\noind ${}^h$ FEST, Schmeilweg 5, D-69118 Heidelberg, Germany \\
\noind ${}^i$ Hiroshima University of Economics, Hiroshima 731-0192, Japan\\
\noind ${}^j$ Yukawa Institute for Theoretical Physics,
              Kyoto University, Japan\\

\begin{center}
{\bf ABSTRACT}
\end{center}
 We study charmonium using the standard relativistic formalism in the quenched
approximation, on a set of lattices with isotropic lattice spacings ranging 
from 0.1 to 0.04 fm. We concentrate on the calculation 
of the hyperfine splitting between $\eta_c$ and $J/\psi$, aiming for a 
controlled continuum extrapolation of this quantity.  
The splitting extracted from the non-perturbatively improved clover Dirac 
operator shows very little dependence on the lattice spacing for $a\leq 0.1$fm. 
The dependence is much stronger for Wilson and tree-level improved clover
operators, but they still yield consistent extrapolations if sufficiently
fine lattices, $a\leq 0.07$ fm ($a M(\eta_c) \leq 1$), are used. 
Our result for the hyperfine splitting is $77(2)(6)$ MeV
(where Sommer's parameter, $r_0$, is used to fix the scale). 
This value remains about 30$\%$ below experiment. 
Dynamical fermions and OZI-forbidden diagrams both contribute to the remainder.
Results for the $\eta_c$ and $J/\psi$ 
wave functions are also presented.
\newpage

\section{Introduction}
\label{s.intro}
Heavy QCD quarkonium systems have been thoroughly studied analytically within
the heavy quark non-relativistic approximation, NRQCD \cite{Lepage}, 
and related heavy quark effective theories as pNRQCD \cite{Pineda} and
vNRQCD \cite{Manohar}-- for a recent review see \cite{nranaly}. 
These approaches, however, generically fail for charmonium.
For $c \bar{c}$ the expansion parameter $v^2\!/\!c^2$ of the effective theory 
is about $0.3$, and higher order corrections 
in $v/c$ seemingly become very large, even overwhelming the leading order
terms. It is particularly challenging to reproduce the hyperfine splitting 
between the $\jpsi$ and the $\etac $ states, which for charmonium is 
$M(J/\psi - \eta_c)=117$ MeV. The  lattice version of NRQCD 
predicts a value of the hyperfine splitting
$M(J/\psi - \eta_c)= \!55(5)$ MeV  \cite{Trottier} far below the 
experimental value,
although still quite remarkable taking into account all the approximations 
involved. Indeed lattice NRQCD, being an effective action with 
cutoff given by the heavy quark mass $m_q$, does not allow for a 
continuum extrapolation: for the effective action to be valid 
one has to preserve $a m_q >1$. Given this, the estimation of the systematic 
uncertainties inherent in the discretization is quite difficult. This fact,
together with the observation that the $v/c$ expansion is not well 
justified for charmonium, encourages the suspicion that the origin of the
discrepancy between experimental and computed hyperfine splitting
could lie in the non-relativistic approximation.
However, all other lattice determinations based on relativistic
actions also underestimate the value of the hyperfine splitting, by as much
as $30\!-\!50\%$ \cite{UKQCD,Klassen,Cppacs1,Cppacs2,Chen}. 
This raises what one might call the puzzle of the hyperfine splitting.

Almost all lattice calculations up to now, including NRQCD, have been 
performed within the quenched (valence quark) approximation (for recent
reviews on heavy quarks on the lattice see \cite{Lproc,McNeile}). 
The effect of quenching has been estimated \cite{Elkhadra} by looking at the 
predicted form of the hyperfine splitting in the heavy quark approximation
\be
M(J/\psi - \eta_c) = \frac{32 \pi \alpha_s(m_q)}{9 m_q^2} |\Psi(0)|^2 \quad ,
\ee
with $\Psi(0)$ the value of the non-relativistic wave function at the origin.
In Ref. \cite{Elkhadra} it is argued that the change in $\Psi(0)$ and in 
the running of $\alpha_s$ due to dynamical 
quarks can give altogether a deviation of the quenched result from the
real world case  as large as $40\%$. 
This would make the hyperfine splitting a quantity particularly
suitable for unveiling unquenching effects -- typically the effects
of quenching on other spectral quantities amount to only $\sim 10\%$.
However,  first numerical results including 
dynamical quarks seem to indicate a much milder $N_f$ dependence of 
the hyperfine splitting than what is needed 
to match the experimental result \cite{Din,El-Khadra:zs}. This raises 
some worry about the reliability of the calculations performed 
up to now, in particular since they have been typically done at not so 
small values of the lattice spacing. This worry becomes more severe after
noting that continuum extrapolations of the hyperfine splitting seem to depend
quite strongly on the kind of Dirac operator used for the calculation
\cite{Trottier,Cppacs2,Din}.
Needless to say that the continuum limit is unique. Such strong dependence
on the choice of Dirac operator has to reflect the existence of large lattice
artifacts.

Indeed, the reason why a reliable determination of the hyperfine splitting 
by means of purely non-perturbative relativistic calculations remains 
elusive is that charm is too heavy for most current lattice simulations. 
For charm the dominant lattice artifacts are of ${\cal O} ((am_c)^n),\ 
n\ge 1$, with $a m_c \sim 0.5$ at the typical values of the lattice spacing.
Lattice artifacts remain large and may completely spoil the determination of
the charmonium spectrum.

One particular approach that has been advocated in order to avoid
large ${\cal O}(am_q)$ lattice artifacts for heavy quark systems
is the use of anisotropic lattices 
\cite{Klassen}. Such lattices \cite{Karsch} have different lattice 
spacings $a_s$ and $a_t$ in spatial and temporal directions, with
$a_t << a_s$. In Ref. \cite{Klassen} it has been argued that by tuning 
appropriately the parameters of the action one could achieve reduced
scaling violations of $O((a_t m_q)^n)$ while keeping still $a_s m_q$ large.
This is certainly a very attractive possibility, which would make anisotropic 
lattices quite advantageous over standard isotropic ones. In particular 
this is the approach that has been taken in several recent studies of the 
charmonium spectrum \cite{Klassen,Cppacs1,Cppacs2,Chen,Matsu2}. 
However, recent analysis seem to point out that scaling violations governed 
by the large $a_s m_q$ artifacts could revive at the one loop level 
\cite{Matsu,Aoki}. This casts doubt on the generic effectiveness of 
anisotropic actions in reducing lattice artifacts better than standard 
isotropic lattices. To settle this point, an analysis of the range of validity 
of the anisotropic approach, as performed in \cite{Matsu3}, seems
essential.

Our methodology here is to compute $M(J/\psi - \eta_c)$ on very fine {\em isotropic} lattices 
within the quenched relativistic formalism. If the lattices used are 
sufficiently fine, this approach should allow for a controlled quenched 
continuum extrapolation, free of the systematic uncertainties which
may have affected other determinations up to now. Particularly important
for this will be the use of a non-perturbatively improved version of the
Dirac operator, for which lattice artifacts are reduced to 
${\cal O} ((am_q)^2)$. Preliminary results of this calculation have 
been presented in \cite{Qcdt}.

\begin{figure}[htb]
\vspace{3.5cm}
\includegraphics{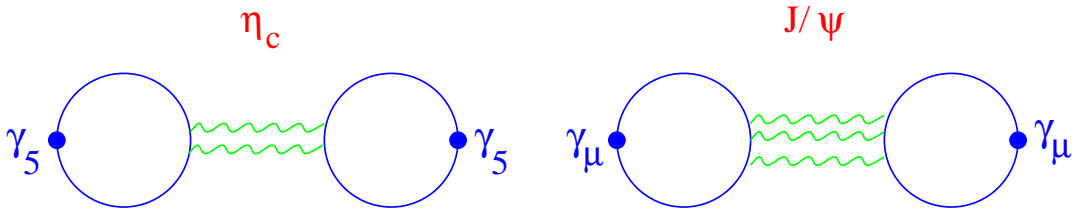}
\caption{ Zweig rule forbidden diagrams contributing to the pseudoscalar
(Left) and vector (Right) channels. These closed quark loops are 
connected by gluonic contributions,  of which we only show those that contribute
to lowest order in $\alpha_s$.}
\label{f:ozi}
\end{figure}

We want to stress that our calculation, which does not differ from any other
in this respect, involves one other approximation 
besides quenching: OZI, meaning that Zweig-rule forbidden diagrams 
are not taken into account\footnote{We thank Stefan Sint for  
this remark.}. Such diagrams, as in Fig.~\ref{f:ozi}, contribute to the 
correlator of flavor singlet mesons like charmonium. They have been 
particularly studied for light quarks in connection with the U(1)$_A$ 
symmetry breaking and the $\eta'$ mass, where the anomaly provides an 
enhancement of OZI contributions to pseudoscalar mesons - for lattice 
estimates see \cite{Singlet1,Singlet2}. There is however no lattice calculation
of their magnitude for heavy quarkonium states like charmonium.
In the quark model \cite{Alvaro,Isgur2} these OZI amplitudes are expected to be
proportional to the value of the wave function at the origin, and hence are
suppressed for P-wave channels which have vanishing wave function at
$\vec r = \vec 0$. For S-wave, their contribution is
${\cal O}(\alpha_s^2 (m_{PS}))$ and ${\cal O}(\alpha_s^3 (m_{V}))$ suppressed
for the pseudo-scalar and vector channels respectively.
Experimental evidence of the goodness of the OZI rule and of this
relative suppression comes from the hadronic widths of heavy pseudo-scalar
and vector mesons. The effect of these diagrams is therefore
expected to be small for heavy quark systems like charmonium. 
Here, we neglect such ``disconnected'' diagrams because of their high computer
cost. We want to stress, however, that their contribution may amount to several
MeV, perhaps a not so negligible effect given the small value of the hyperfine 
splitting. Moreover, the OZI contribution could be enhanced by mixing with 
glueballs with the appropriate quantum numbers provided these glueballs were 
almost degenerate in mass with the charmonium states, an effect that has been 
measured for light singlet mesons \cite{Singlet1,Singlet3}.

The paper is organized as follows. Section 2 compiles our results. 
In Section 2.1  details about our numerical simulations are presented,
including the kind of Dirac operators used to measure meson masses and the 
strategy to extract the hyperfine splitting from 2-point correlators.
In Section 2.2 we give our values for the $\hypsp$ hyperfine splitting 
as well as the $1P$ to $\jpsi$ splittings, obtained  with the 
non-perturbatively improved clover Dirac operator. Our continuum extrapolation
of the charmonium spectrum is based on data at lattice spacings ranging from
$a= 0.0931$ fm to  0.0397 fm. A comparison with the 
results obtained  from Wilson and tree-level clover Dirac operators is 
given in Section 2.3. In Section 2.4 we present the results for the 
$\eta_c$ and $J/\psi$ wave functions. We investigate the volume 
dependence or our results in Section 2.5, by looking at the dependence 
on physical volume of both the spectrum and the wave functions. Finally, our
summary and conclusions are presented in Section 3.

\section{Results}

We have aimed at analyzing the S- and P-wave states indicated in 
Table \ref{t:states}.
Precise results will only be presented for the $\hypsp $ hyperfine 
splitting, although preliminary results for P-wave meson masses will also be 
given. 

\begin{table}[htb]
\caption{ Charmonium S- and P-wave states analyzed in this work. We indicate
their quantum numbers and physical masses. Unless indicated experimental errors
are one or less in the last indicated digit.}
\vspace*{0.1cm}
\label{t:states}
\newcommand{\m}{\hphantom{$-$}}
\newcommand{\cc}[1]{\multicolumn{1}{c}{#1}}
\renewcommand{\tabcolsep}{2.889pc} 
\renewcommand{\arraystretch}{1.2} 
\begin{tabular}{@{}|l|lll|}
\hline
Name&$^{(2s+1)}L_J $&$J^{PC}$ & Mass(GeV)\\
\hline
\m$\eta_c$   & \m$^1 S_0$& $0^{-+}$& \m 2.980(2) \\
\m$J/\psi$   & \m$^3 S_1$& $1^{--}$& \m 3.097 \\
\m$h_c$      & \m$^1 P_1$& $1^{+-}$& \m 3.526 \\
\m$\chi_{c0}$& \m$^3 P_0$& $0^{++}$& \m 3.415 \\
\m$\chi_{c1}$& \m$^3 P_1$& $1^{++}$& \m 3.511 \\
\hline
\end{tabular}\\[2pt]
\end{table}
 
\subsection{Simulation details}

Gauge configurations are generated with the standard Wilson action. Simulation
parameters - see Table \ref{t:simp} - have been chosen so as to fix a
physical lattice size of about 1.6 fm. In addition we have also simulated
a very fine $32^3\times 96$ lattice at $\beta=6.6$, corresponding to a
smaller size of 1.3 fm (possible finite volume effects will be discussed
in section \ref{s:fv}). The physical scale is set by $r_0$ from \cite{Sommer}.

\begin{table}[htb]
\caption{Simulation parameters. The scale is set by $r_0$ from  \cite{Sommer}.
The non-perturbative value of the clover coefficient, $c_{sw}^{\rm NP}$,
has been computed in \cite{Alpha}.}
\vspace*{0.1cm}
\label{t:simp}
\newcommand{\m}{\hphantom{$-$}}
\newcommand{\cc}[1]{\multicolumn{1}{c}{#1}}
\renewcommand{\tabcolsep}{1.6pc} 
\renewcommand{\arraystretch}{1.2} 
\begin{tabular}{@{}|llllll|}
\hline
$\beta$&$L^3\times T$ &  $a ({\rm fm})$ & $L a ({\rm fm})$
&$c_{sw}^{\rm NP}$&$\#$ conf.  \\
\hline
6.0  & $18^3\times 48$    & 0.0931 & 1.68 &1.769  & \m190\\
6.2  & $24^3\times 72$    & 0.0677 & 1.62 &1.614  & \m  90\\
6.4  & $32^3\times 96$    & 0.0513 & 1.64 &1.526  & \m  60\\
6.6  & $32^3\times 96$    & 0.0397 & 1.27 &1.467  & \m 130\\
\hline
\end{tabular}\\[2pt]
\end{table}

Quark propagators are computed using Wilson, tree-level clover and 
non- perturbatively improved clover Dirac operators. The non-perturbative value
of the clover coefficient is taken from \cite{Alpha}.

A complete analysis of the spectrum, at all $\beta$ values, has only 
been performed for the non-perturbatively improved clover Dirac operator. 
For Wilson and tree-level clover Dirac operators we have results 
at $\beta=6.2$ (for Wilson), $\beta=6.4$ and $\beta=6.6$. 
The continuum extrapolations 
in such cases are based on available data at $\beta\!=\!6.0$ and 6.2  
from UKQCD \cite{UKQCD,Collins}.  

\begin{figure}[htb]
\vspace{9.0cm}
\includegraphics{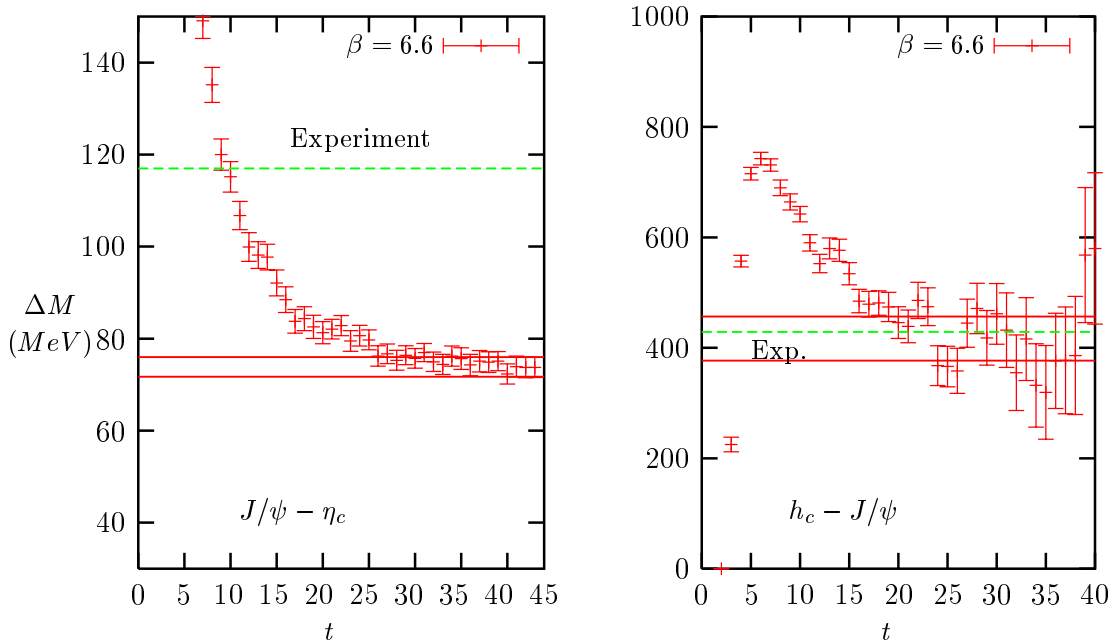}
\caption{Left: $\hypsp$ hyperfine splitting effective mass.
Right: $ ^1P_1- \jpsi$ splitting effective mass. Data are obtained using
the non-perturbatively improved clover Dirac operator. The dashed horizontal 
line is the experimental value and the band comprises
our fit to a constant with its errors.}
\label{fig1}
\end{figure}

We compute correlation functions of hadronic operators with zero-momentum
point source and point sink. A time dependent effective mass 
$M(t)$ is extracted by fitting the ratio of correlators at 
a pair of points ($t$, $t+1$) to:
\be
R(t)\equiv \frac{\<C(t)\>}{\<C(t+1)\>} = \frac{{\rm cosh} 
[(\frac{T}{2}-t) \ M(t)]}{{\rm cosh} [
(\frac{T}{2}-t-1) \ M(t)]}.
\label{eq:efm}
\ee
Here $\langle .\rangle$ means summation over Yang-Mills configurations.
The plateau in the effective mass is then fitted to a constant which provides
our estimate of the meson mass. 

In addition, we extract mass differences between the
states in Table \ref{t:states}  from the ratio of correlators, 
which behaves as:
\be
\frac{R_A(t)}{R_B(t)} = \frac{{\rm cosh} 
[(\frac{T}{2}-t)  M_A(t)] \ 
{\rm cosh}
[(\frac{T}{2}-t-1)  M_B(t)]}
{{\rm cosh}
[(\frac{T}{2}-t)  M_B(t)] \ 
{\rm cosh} [(\frac{T}{2}-t-1)  M_A(t)]}
\ \ \sim_{_{_{\!\!\!\!\! \!\!\!\!\!\!\!T\rightarrow \infty}}} 
e^{-\Delta M(t)}\quad ,
\label{eq:split}
\ee
if the correlators are dominated by a single pole.
This allows to extract a time dependent effective mass for 
the splitting, $\Delta M(t) \equiv M_B(t)
 - M_A(t)$, which is fitted to a constant
in the range where the effective mass plateau sets in. The quality of the 
data for the hyperfine splitting is shown in Fig.~\ref{fig1} left.  
A similar plot for the $^1P_1- \jpsi $ splitting is presented in 
Fig.~\ref{fig1} right. Although the signal extracted from Eq.~(\ref{eq:split})
is less noisy than the one from Eq.~(\ref{eq:efm}), point sources are still too 
noisy for P-wave states and, within our statistics, do not allow a 
precise determination of the masses. 

\subsection{Charmonium spectrum from non-perturbatively \\ improved
clover Dirac operator}

We present first the results obtained with the
non-perturbatively improved clover Dirac operator, for which scaling
violations in spectral quantities are expected to be ${\cal O}(a^2)$.
A detailed comparison with Wilson and tree-level clover Dirac operators will
be presented in section \ref{s:comp}.

\begin{table}[htb]
\caption{Charmonium spectrum from non-perturbatively
improved clover Dirac operator. Results are given in MeV
with the scale set by $r_0$ from  \cite{Sommer}.
$\kappa$ has been tuned to maintain an approximately  constant mass
$M(\jpsi) \!\approx \!3095$ MeV $\forall a$.
The last two columns show our continuum extrapolation
and the experimental value.}
\vspace*{0.1cm}
\label{t.clnp2}
\newcommand{\m}{\hphantom{$-$}}
\newcommand{\cc}[1]{\multicolumn{1}{c}{#1}}
\renewcommand{\tabcolsep}{0.33pc} 
\renewcommand{\arraystretch}{1.2} 
\begin{tabular}{@{}|l|llll|ll|}
\hline
&\m$\beta=6.0$ & \m$\beta=6.2$ & \m$\beta=6.4$&\m$\beta=6.6$&
\m$a\rightarrow 0$& Exp.\\
&$\kappa=0.11865$&$\kappa=0.12457$&$\kappa=0.12755$&
$\kappa=0.12943$&&\\
\hline
$\etac$ &\m3023(2)&\m3019(3)&\m3034(3)&\m3014(3)&& 2980\\
$\jpsi$ &\m3091(2)&\m3093(3)&\m3109(4)&\m3085(3)&& 3097\\
$\hypsp$&\m68.5(1.3)&\m74.2(1.5)&\m75.2(2.5)&\m73.8 (2.1)&\m 77.2(1.7) &$\ 117$\\
$ ^1P_1-\jpsi$&\m417(25)&\m460(34)&\m433(37)&\m417(40)&\m441(25)&\ 429   \\
$ ^3P_0-\jpsi$&\m342(19)&\m352(25)&\m369(34)&\m397(41)&\m387(14)&\ 318   \\
$ ^3P_1-\jpsi$&\m390(19)&\m413(34)&\m451(41)&\m417(29)&\m437(16)&\ 414   \\
\hline
\end{tabular}\\[2pt]
\end{table}

We have used two different ways of setting the charm quark point:
imposing that either $r_0 M(\etac) $ or $r_0 M(\jpsi)$ equals
the physical value. These two different choices allow to study the
ambiguities in the scale determination induced by the approximations we
have made: quenched and OZI.
We want to stress again that our calculation is affected not only by
quenching ambiguities but also by OZI effects, i.e. only
a subset of the relevant quenched diagrams has been included.
Taking this into account we expect the choice of $J/\psi$ mass as
reference to be a better one, since in that case OZI contributions
to the $J/\psi$ mass are expected to be considerably suppressed compared to 
the pseudoscalar ones - see for instance \cite{Isgur} and the discussion in
Section \ref{s.intro}. 
For these reason the numbers presented here refer to the choice of $J/\psi$ 
mass as reference. The hyperfine splitting with $M(\etac)$
fixed to the experimental $\eta_c$ value will
be used below when we estimate the systematic uncertainty in our calculation. 
An alternative possibility, which we have not explored, would be to fix the 
charm quark point by using the $D_s$ meson mass, as done for instance in 
heavy-light spectroscopy and for the determination of the charm quark mass 
\cite{Stefan,UKQCD2}. This would be free of OZI ambiguities. 

\begin{figure}[hp]
\vspace{8.0cm}
\includegraphics{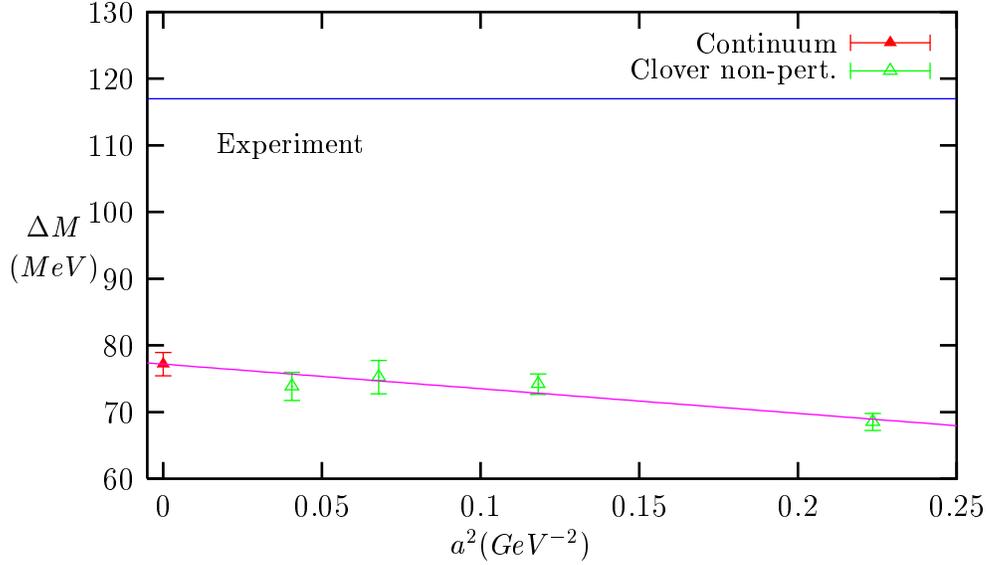}
\caption{Continuum extrapolation of the hyperfine splitting with the
non-perturbatively improved clover Dirac operator. The bare quark mass
is tuned to maintain an approximately constant mass $M(\jpsi) \approx 3095$ MeV $\forall a$.}
\label{f:cswhyp}
\end{figure}

Our results are collected in Table \ref{t.clnp2}. 
As indicated above, the charm quark point has been
obtained by fixing $M(\jpsi) \approx 3095$ MeV for all $a$.
We compute the hyperfine splitting from: 
$(A)$ the difference between the $\jpsi$ and $\etac$ masses and $(B)$ the
hyperfine splitting effective mass extracted from the ratio of correlators
Eq.~(\ref{eq:split}). Both determinations are perfectly consistent within
errors but (B) is more precise and will be used henceforth.
Fig.~\ref{f:cswhyp} shows the results for the $\hypsp$ hyperfine splitting as
a function of $a^2$.  The lattice spacing dependence is very small.  
On the coarsest lattice, for which $aM({\etac})\approx 1.4$ and large scaling 
violations could be expected, the deviation from our continuum 
extrapolation amounts to only $11\%$.  
The cutoff dependence is well fitted linearly in $a^2$.
The continuum extrapolation is $ M({\hypsp})=77(2)$ MeV.
Excluding from the fit the point at $\beta=6.0$ gives 
$M({\hypsp})=74(1)$ MeV. If, instead of the vector mass, we  
fix the charm scale by setting  $M(\etac) \approx 2.945$ MeV $\forall a$, 
the result for the hyperfine splitting goes up by $6\%$. Including both these 
results as systematic error in our 
determination we quote as value of the hyperfine splitting from the 
non-perturbatively improved clover Dirac operator $ M({\hypsp})=77(2)(6)$.  
This is about $30\%$ below the experimental value. 

When comparing our number to previous lattice determinations - 
see \cite{Cppacs1,Cppacs2,UKQCD,Trottier} - it is important to note that 
we have used $r_0$ to set the scale. It is is quite common for charmonium 
analysis to use instead the spin averaged splitting, $\ ^1P_1-\avs$ or 
$\avp-\avs$, which is claimed to lead to considerably larger values of the 
hyperfine splitting. Variations with the scale input are usually blamed on 
the quenched approximation \footnote{Note, however, that this quantity is also 
affected by the OZI  approximation}. Although this is partially true,
the large differences often quoted are also coming from large scaling 
violations. This has been already illustrated by the latest CP-PACS result  
\cite{Cppacs2}, obtained using 
the Fermilab anisotropic action. The discrepancy in the determination of the
hyperfine splitting from $r_0$ or $\avp-\avs$ has been
reduced from 27 to 16$\%$ as the lattices used in the continuum extrapolation
have changed from $a\in [0.099, 0.208]$ fm  \cite{Cppacs1} 
to $a\in [0.0697,0.1374]$ fm \cite{Cppacs2}. Their final number is
$M({\hypsp })_{\avp-\avs}=85.3(4.4)^{(+5.7)}_{(-2.5)}$ MeV. From $r_0$ they 
obtain instead $M({\hypsp })_{r_0}= 72.6(9)^{(+1.2)}_{(-3.8)}$ MeV (the 
previous result in \cite{Cppacs1} was $M({\hypsp })_{r_0}= 65(1)$ MeV, the 
discrepancy clearly reflecting the large systematic ambiguities due to scaling 
violations). Our final number, $M({\hypsp })_{r_0}= 77(2)(6)$, lies in 
between their two determinations, and is compatible within errors with both.

\begin{figure}[hp]
\vspace{8cm}
\includegraphics{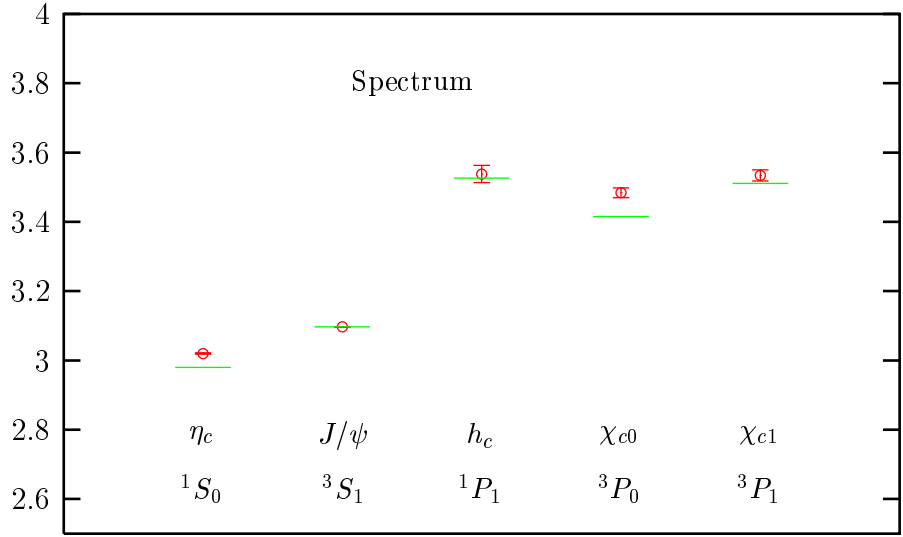}
\caption{Comparison between the experimental charmonium spectrum and
our numerical data. The scale is set by $r_0$. The value of the $J/\psi$ mass
is fixed to the experimental one.}
\label{f:spect}
\end{figure}

We have also computed the splitting between P-wave 
states and the $ \jpsi$. Results are presented in Table \ref{t.clnp2}. 
With point sources and within our limited statistics, our results are too 
noisy to attempt a reasonable continuum extrapolation and make a definite 
statement about scaling violations in these quantities. 
Still, our estimates of the $J=1$ to $J/\psi$
splittings are consistent, within errors, with experiment. The discrepancy 
turns out to be larger for the $^3 P_0$ state.

An estimate of the ambiguities inherent in the quenched plus OZI approximations
can be extracted from the values of the continuum extrapolated spectrum. Figure 
\ref{f:spect} presents a comparison between the experimental charmonium spectrum
and our data. Deviations from experiment amount to $30\%$ for the $J/\psi$
to $\eta_c$ splitting, and $22\%$ for the $^3P_0$ to $J/\psi$  
splitting. They are considerably reduced for the $J=1$, P-wave to $J/\psi$ 
splittings, which are consistent within errors with experiment.

\subsection{Dependence on the choice of Dirac operator and
continuum extrapolations.}
\label{s:comp}

As discussed in the Introduction, it has been often reported 
\cite{Trottier,Cppacs2,Din} that the continuum 
extrapolated hyperfine splitting strongly depends on the choice of Dirac 
operator. This happens in particular with anisotropic actions like the 
ones used in \cite{Klassen,Cppacs1,Cppacs2} which are claimed to reduce 
${\cal O} (m_q a_s)^n$ scaling violations to ${\cal O} (m_q a_t)^n, 
\ a_t \ll a_s$.
Since the continuum limit is unique, such strong dependence 
on the choice of Dirac operator necessarily reflects the existence of 
large lattice artifacts. Indeed,  CP-PACS, in Ref. \cite{Cppacs2}, concludes
that $ m_q a_s< 1$ is still certainly 
needed in order to obtain reliable continuum extrapolations with
anisotropic lattices. This is not surprising,
but it is an indication that anisotropic lattices 
might not, in general, succeed in removing scaling violations better than
isotropic lattices do (indications that ${\cal O} (g^2 m_q a_s)$ indeed revive 
at the one-loop level have been reported in \cite{Matsu,Aoki}). As we will see 
next, even with the
non-improved Wilson Dirac operator a reasonable estimate of the hyperfine
splitting can be obtained if, but only if, lattices with spacing $a\le 0.07$ fm,
i.e. $aM(\etac) \le 1$, are used.

\begin{table}[htb]
\caption{Pseudoscalar, vector mass and hyperfine splitting from Wilson and
tree-level clover Dirac operator. Results are given in physical
units (MeV) with the scale set by $r_0$.}
\vspace*{0.2cm}
\label{t:noimp}
\newcommand{\m}{\hphantom{$-$}}
\newcommand{\cc}[1]{\multicolumn{1}{c}{#1}}
\renewcommand{\tabcolsep}{1.3pc} 
\renewcommand{\arraystretch}{1.2} 
\begin{tabular}{@{}|l|ll|lll|}
\hline
Dirac Op. &$\kappa$&$\beta$& $\etac$&$\jpsi$&$\hypsp$\\
\hline
Wilson    &0.1380       &  6.2       &2728(3)   & 2764(4)&35.2(1.2) \\
Wilson    &0.1375       &  6.2       &2788(3)   & 2821(3)&33.2(0.9) \\
Wilson    &0.1365       &  6.2       &2913(3)   & 2943(3)&30.4(0.9) \\
Wilson    &0.1350       &  6.2       &3099(4)   & 3125(4)&26.7(0.9) \\
\hline
Wilson    &0.1389       &  6.4       &2925(4)   & 2963(4)&38.4 (1.6) \\
Wilson    &0.1380       &  6.4       &3087(4)   & 3122(5)&35.4 (1.7) \\
Wilson    &0.1371       &  6.4       &3263(5)   & 3288(4)&28.5 (1.1) \\
\hline
Wilson    &0.1415       &  6.6       &2594(5)   & 2652(6)&60.4 (2.4) \\
Wilson    &0.1400       &  6.6       &2970(6)   & 3017(7)&48.3 (2.3) \\
Wilson    &0.1385       &  6.6       &3338(5)   & 3375(5)&39.1 (1.6) \\
Wilson    &0.1375       &  6.6       &3583(5)   & 3613(5)&32.3 (1.6) \\
\hline
Clover    &0.1324       &  6.4       &2931(4)   & 2996(5)&64.5 (2.8) \\
Clover    &0.1320       &  6.4       &3022(4)   & 3081(4)&59.2 (2.5) \\
Clover    &0.1315       &  6.4       &3129(4)   & 3189(5)&59.0 (2.4)  \\
\hline
Clover    &0.1335       &  6.6       &2964(6)   & 3031(6)&68.9 (3.2) \\
Clover    &0.1330       &  6.6       &3116(6)   & 3180(6)&64.8 (3.0) \\
Clover    &0.13225      &  6.6       &3342(6)   & 3399(6)&58.9 (2.8) \\
\hline
\end{tabular}\\[2pt]
\end{table}

The pseudoscalar mass and the hyperfine splitting from
Wilson and tree-level clover Dirac operators are presented in
Table \ref{t:noimp}. We only have data at $\beta=6.2$ (for Wilson),
$\beta=6.4$ and $\beta=6.6$.
To perform the continuum extrapolation we use $\beta\!=\!6.0$ and 6.2
data from UKQCD \cite{UKQCD,Collins}. Our strategy here has been
slightly different than in the previous section. Instead of fine
tuning the vector mass at each $\beta$ value, 
we interpolate from a range of masses around the physical $J/\psi$ mass.
We fit the dependence of $M(\hypsp)$ versus $1/M(\jpsi)$ at fixed 
$\beta$, and extract the value of the hyperfine splitting at the desired 
vector meson mass from the fit.  
This allows to extract the splitting at $M(\jpsi)=3.095$ GeV, the
same vector meson mass which we fixed for the non-perturbatively improved Dirac 
operator.

Concerning the continuum extrapolations we expect scaling violations to be:
\begin{itemize}
\item  ${\cal O}(a)$ for the Wilson Dirac operator.
\item  ${\cal O}(a^2)$ and ${\cal O}(g^2 a)$  for the tree-level clover  
Dirac operator.
\end{itemize}
These are the dominant contributions but, for very coarse lattices, 
sub-leading terms might also be important especially since they depend on $(am_q)^2$. To test the approach to the
continuum, we present in Fig. \ref{f:extcomp1} the results of two different
extrapolations:

\begin{enumerate}
\item [(I)] Top figure \ref{f:extcomp1}: using the three largest $\beta$ values,
\begin{itemize}
\item linearly in $a$ for the Wilson Dirac operator
\item linearly in $a^2$ for the tree-level clover Dirac operator
\end{itemize}

\begin{figure}[hp]
\vspace{15.0cm}
\includegraphics{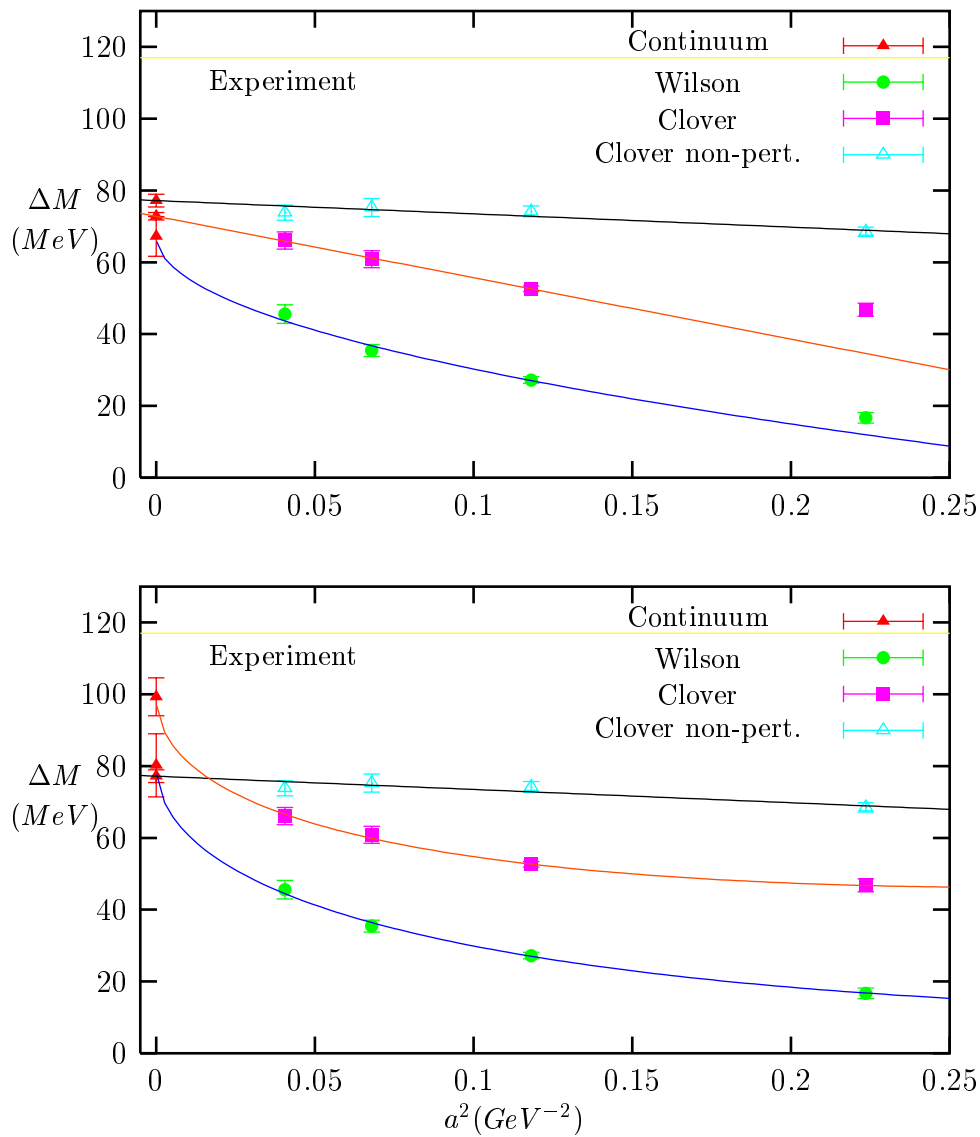}
\caption{Comparison between continuum extrapolation of $M(\hypsp)$ with
Wilson, tree-level clover and  non-perturbatively improved clover
Dirac operators. Top: Wilson  and tree-level clover improved data are fitted
linearly in $a$ and $a^2$ respectively.
Bottom: Both Wilson and tree-level clover data are fitted
with $a$ plus $a^2$ dependence. Wilson data at $\beta=6.0$ and 
tree level clover data at $\beta=6.0$ and $\beta=6.2$ are based on UKQCD
results in \cite{UKQCD,Collins}.}
\label{f:extcomp1}
\end{figure}

\item [(II)] Bottom figure \ref{f:extcomp1}: using all four $\beta$ values,
\begin{itemize}
\item including $a$ and $a^2$ terms for the Wilson Dirac operator
\item including $a$ and  $a^2$ terms for the tree-level clover Dirac operator
\end{itemize}
\end{enumerate}

 We obtain as continuum extrapolations for the Wilson and tree-level clover
respectively:
$M(\hypsp)=67(6)$ MeV and $73(1)$ MeV  from fit (I) and  $M(\hypsp)= 80(9)$ MeV
and $99(5)$ MeV from fit (II). This is to be compared with the result from
non-perturbative improvement $ M({\hypsp})=77(2)$ MeV.

Several remarks are in order here: 

$(i)$ Only the non-perturbatively improved data show a weak dependence
on the lattice spacing.
Wilson and tree-level clover data are both significantly below their continuum
extrapolations at all the simulated $\beta$ values. 

$(ii)$ Linear extrapolations in $a$ and $a^2$ for Wilson and tree-level 
clover respectively, including $\beta\!=\!6.0$ ($aM({\etac})=1.4$) are not 
justified. 


$(iii)$ If too coarse lattices are included in the fit, continuum 
extrapolations become quite sensitive to the assumed
dependence on the lattice spacing. This is clearly observed
in the case of the tree-level clover data, for which  fit (II) gives a 
considerably higher continuum value.  It is only when the extrapolations
start from sufficiently fine lattices that they come out reasonably 
consistent. A lower bound for consistent extrapolations seems to be
$a\leq 0.07$ fm for Wilson and tree-level clover data, i.e. $a M(\etac)\leq 1$. 
Surprisingly, though,
non-perturbative clover improvement seems to work well even on the
coarser $\beta=6.0$ lattice where $aM({\etac})=1.4$.

To summarize, non-perturbative clover improvement seems crucial to remove 
strong scaling violations. It is possible to extract comparable results from 
Wilson or tree-level clover improved Dirac operators, but very fine 
lattices are needed in order to remove the ambiguities inherent in the 
extrapolation procedure.

\subsection{Wave functions}

The origin of the the hyperfine splitting can be easily understood
within the naive non-relativistic approximation (see for instance \cite{Ynd}). 
This approximation amounts to solving the Schr\"odinger equation in a 
non-relativistic Coulombic potential and dealing with relativistic 
corrections in perturbation theory. To zeroth order $^3 S_1$ and $^1 S_0$ 
states are degenerate. The degeneracy is removed to first order in 
perturbation theory by the spin-spin interaction, giving a value of 
the hyperfine splitting:
\be
M(\hypsp)= \frac{32 \pi \alpha_s(m_q)}{9 m_q^2} |\Psi_{\rm NR}(0)|^2
\ee
with $\Psi_{\rm NR} (0)$ the value of the non-relativistic wave function 
at the origin ($\Psi_{\rm NR}(r)=  (8 \pi \rho^3)^{-1/2} \exp\{-r/(2\rho)\}$,
$\rho=(4 \alpha_s m_q/3)^{-1}$).
Perturbative corrections to the wave function also depend on the spin; 
to lowest order in perturbation theory, the value of the wave function at 
the origin increases for the pseudoscalar and decreases for the vector 
according to \cite{Ynd,Ynd2}:
\bea
\Psi_{\eta_c}(0) = \Big (1+\delta_{\rm NP}+(\frac{1}{2} -\nu) 
\frac{8\ \alpha_s^2(\mu^2)}{9}\Big) \Psi_{\rm NR}(0)\, \label{nr1}\\
\Psi_{J/\psi}(0) =\Big(1+\delta_{\rm NP}-(\frac{1}{6} +\nu) 
\frac{8\ \alpha_s^2(\mu^2)}{9}\Big)\Psi_{\rm NR}(0)\label{nr2}.
\eea
where $\nu \approx 7.241 \times 10^{-2}$ and $\alpha_s(\mu^2)$ the 
strong running coupling evaluated at scale $\mu^2$. Here
$\delta_{\rm NP}$ denotes the, spin-independent, non-perturbative correction 
to the wave function at the origin (estimated in \cite{Ynd2,Leutw}).  

\begin{figure}[hp]
\vspace{9.0cm}
\includegraphics{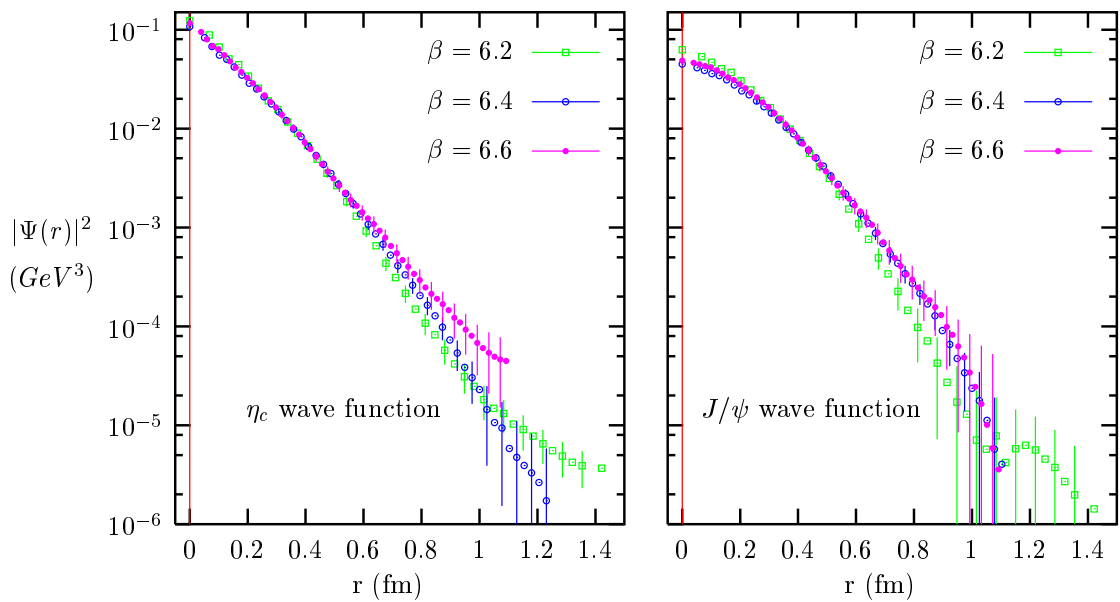}
\caption{Scaling analysis of pseudoscalar (left) and vector (right) matter
wave functions. The vertical scale is logarithmic.
}
\label{fig.wavef}
\end{figure}

As in Ref. \cite{Alexandrou}, we have extracted gauge invariant wave 
functions from lattice 4-point functions: 
\bea
 |\Psi (\vec{r})|^2&=& \langle c \bar{c} | (\psi_c^\dagger \psi_c)(0) 
(\psi_{\bar{c}}^\dagger \psi_{\bar{c}})(\vec{r}) | c \bar{c} \rangle = 
\label{eq:wavef}\\
 {\cal N} \sum_{\vec x } 
\langle \ {\rm Tr} &\{ & S(\vec 0,0; \vec{x}, t_1)\  Q\  S(\vec 0,2 t_1; \vec{x}, t_1)^\dagger 
\Gamma \gamma_5 \nonumber \\
&& S(\vec 0,2 t_1; \vec{x}+\vec{r}, t_1)  \ Q^\dag \ S(\vec 0,0; \vec{x}+\vec{r}, t_1)^\dagger 
\gamma_5 \Gamma^\dagger \} \ \rangle \nonumber
\eea
$S$ is the quark propagator and $\Gamma = \{\gamma_5,  \gamma_i\}$ 
for $\{{\rm \etac, \jpsi}\}$. We have measured matter and charge wave functions
by setting $Q=\{ 1, \gamma_0\}$ respectively \cite{Michael}. 
${\cal N}$ is a normalization constant fixed by imposing normalization 1 to the 
wave function. 

\begin{table}[htb]
\caption{ Parameters of the fit to Eq. (\ref{eq.fitwf}) of the $\eta_c$ 
and $J/\psi$ matter wave function. We give both the direct
values of the wave function at the origin $\Psi(0)$ and the values derived from
the fit, $\hat\Psi(0)$, in GeV$^3$. 
The parameter $\rho $ is given in fermi.}
\vspace*{0.2cm}
\label{t:fitwf}
\newcommand{\m}{\hphantom{$-$}}
\newcommand{\cc}[1]{\multicolumn{1}{c}{#1}}
\renewcommand{\tabcolsep}{0.28pc} 
\renewcommand{\arraystretch}{1.2} 
\begin{tabular}{@{}|l|llll|llll|}
\hline
$\beta$&$|\Psi_{\eta_c}(0)|^2$&$|\hat\Psi_{\eta_c}(0)|^2$&$p_{
\eta_c}$&$\rho_{\eta_c}$
&$|\Psi_{J/\psi}(0)|^2$&$|\hat\Psi_{J/\psi}(0)|^2$&$p_{J/\psi}
$&$\rho_{J/\psi}$\\
\hline
6.0&0.156(6)&0.153(6) &1.24 (2)& 0.155(4)&0.093(4)&0.089(3) &1.46(3)& 0.210(5)\\
6.2&0.124(7)&0.117(3) &1.24 (2)& 0.172(4)&0.063(3)&0.060(1) &1.59(2)& 0.255(2)\\
6.4&0.108(5)&0.100(2) &1.20 (2)& 0.176(3)&0.045(3)&0.0445(3)&1.59(1)& 0.283(1)\\
6.6&0.117(9)&0.112(2) &1.14 (1)& 0.164(2)&0.049(3)&0.0500(5)&1.57(2)& 0.273(2)\\
\hline
\end{tabular}\\[2pt]
\end{table}

A comparison between pseudoscalar and vector matter wave functions
using the non-perturbatively improved clover Dirac operator
is presented in Fig.~\ref{fig.wavef}. We have binned the wave function in 
bins of size $a/2$. We include in the figure results 
for $\beta=6.2$, 6.4 and $\beta=6.6$ which show very 
good scaling within errors, up to  possible, small, 
finite size effects  for $\beta=6.6$ which will be discussed in the next 
section. The observed pattern corroborates qualitatively the predictions of the 
heavy-quark model: the value of the wave function at the origin increases 
for the pseudoscalar, decreases for the vector.  
Table \ref{t:fitwf} presents the results of the fit 
of matter wave functions ($Q$ equal 1 in Eq.~(\ref{eq:wavef}))
to the following ansatz motivated by the heavy 
quark non-relativistic approximation and the form of 
variational wave functions in potential models:
\be
|\Psi(r)|^ 2=|\hat\Psi(0)|^ 2 \exp\Big\{-\Big(\frac{r}{\rho}\Big)^p\Big\} \,.
\label{eq.fitwf}
\ee
Our fits have been performed for $r\le 0.8$ fm. In all cases
we obtain a $\chi^2/ndf < 0.5$ ($<$1.4 for $\beta=6.0$).
In infinite volume, wave function normalization fixes 
the value of the wave function at the origin to $|\hat\Psi(0)|^2= p/ 
(4\pi \rho^ 3 \Gamma[\frac{3}{p}])$, a relation well satisfied by our fits.
Given the good scaling properties of our $\beta=6.6$ and 6.4 wave functions 
one can extract an estimate of the continuum values of the pseudoscalar and
vector wave functions at the origin. Fitting to a constant the results for 
these two values of $\beta$ we obtain $|\Psi_{\eta_c}(0)|^2=0.110(4)$ and
$|\Psi_{J/\psi}(0)|^2=0.047(2)$. This quantity is of phenomenological
interest since it enters in many of the estimates
of the heavy quark approximation as well as in potential models for heavy 
quarks. One example, apart from the hyperfine splitting, is
the leptonic decay width of the $J/\psi $ which, in the non-relativistic
approximation, can be expressed as: 
\be
\Gamma (J/\psi \rightarrow e^+ e^-) = \frac{16 \pi Q_c^2 \alpha^2}
{m^2_{J/\psi}} |\Psi(0)|^2 
\label{decay}
\ee   
with $Q_c$ the charm quark charge in units of the proton charge
and $\alpha$ the fine structure constant. Inserting our value of the $J/\psi$
wave function at the origin we obtain $\Gamma (J/\psi \rightarrow e^+ e^-) =
5.8 \pm 0.2 $ keV, probably in too good agreement with the experimental value
$\Gamma_{\rm exp} (J/\psi \rightarrow e^+ e^-) = 5.26\pm 0.37$ keV, taking
into account that formula (\ref{decay}) does not include radiative corrections
(appart from those affecting the wave function) which, estimated to lowest 
order, multiply the right hand side of Eq.  (\ref{decay}) by a factor 
$1-16 \ \alpha_s(4 m^2_c) /(3\pi)\sim 0.5 $ \cite{Ynd,Barbieri}.

We can also make use of formulas (\ref{nr1}) and (\ref{nr2}) to extract an 
estimate of the magnitude of non-perturbative contributions to the wave
function at the origin ($\delta_{\rm NP}$).
For this we use $m_c(m_c)=1.301(34)$ GeV from \cite{Stefan} and extract
from the equations both $\delta_{\rm NP}$ and $\alpha_s$.
Plugging our values for $|\Psi_{\eta_c}(0)|^2$ and $|\Psi_{J/\psi}(0)|^2$ 
gives $\delta_{\rm NP}= -0.12$ and $\alpha_s= 0.761$. The 
large value of $\alpha_s$ needed to match our results is
a clear indication that spin dependent, and hence relativistic, 
non-perturbative effects are indeed rather strong for these charmonium states.
If we would instead fix the strong coupling constant to $\alpha_s\sim 0.5$ 
as in \cite{Ynd} we would obtain a strong spin dependence of $\delta_{\rm NP}$
which would moreover turn out to be of ${\cal O}(1)$ for the $\eta_c$ wave 
function.

\subsection{Finite volume effects}
\label{s:fv}

\begin{table}[htb]
\caption{Pseudoscalar mass and hyperfine splitting from non-perturbatively
improved clover Dirac operator. The lattice spacing is fixed to 0.093 fm
($\beta=6.0$) and the number of lattice points $L$, hence the physical 
volume $La$, is varied as indicated in the table. Results, averaged over 
100 configurations (190 for $L=8$), are given in physical units (MeV) 
with the scale set by $r_0$.}
\vspace*{0.1cm}
\label{t:fvol}
\newcommand{\m}{\hphantom{$-$}}
\newcommand{\cc}[1]{\multicolumn{1}{c}{#1}}
\renewcommand{\tabcolsep}{1.6pc} 
\renewcommand{\arraystretch}{1.2} 
\begin{tabular}{@{}|ll|lll|}
\hline
\m$L$ &$La$ (fm)& \m$\etac$& \m$\jpsi$&\m$\hypsp$\\
\hline
\m8  &\m0.75 &\m2958(10)&\m3019(12)&\m61.4(4.4)\\
\m10 &\m0.93 &\m2953(5) &\m3023(6)&\m70.6(2.5)\\
\m12 &\m1.12 &\m2957(3) &\m3032(5)&\m75.4(2.7)\\
\m14 &\m1.30 &\m2947(3) &\m3020(4)&\m72.6(1.9)\\
\m16 &\m1.49 &\m2952(3) &\m3025(4)&\m74.9(2.1)\\
\m18 &\m1.68 &\m2949(2) &\m3021(3)&\m72.5(1.5)\\
\hline
\end{tabular}\\[2pt]
\end{table}

In this section we investigate how our results depend on the physical lattice
volume. In particular we are interested in what happens with our finest
$\beta=6.6$ lattice which has a somewhat small physical size
$La=1.3$ fm. An indication of the magnitude of finite volume effects can
already be obtained from the wave function plots and fits in the previous
section.
Compare in Fig. \ref{fig.wavef}  the results for $\beta=6.6$ ($La$=1.3 fm)
and $\beta=6.4$ ($La$=1.6 fm).  Within errors, the wave functions
do not show any clear signal of finite size effects, except for a small
deviation of the pseudoscalar wave function at the tail, which is, however,
not very significant within our errors. One could also argue from the fits in
Table \ref{t:fitwf} that the wave functions at the origin
are slightly larger than expected for the $La$=1.3 fm lattice, but this
may be just a reflection, through the wave function normalization, of the
finite volume effects at the tail.
It is, anyway, clear from these plots that finite size effects on the
charmonium wave function are really small on the $La$=1.3 fm lattice.
In consequence, we do not expect the hyperfine splitting on this lattice to
be significantly affected.

\begin{figure}[htb]
\vspace*{8cm}
\includegraphics{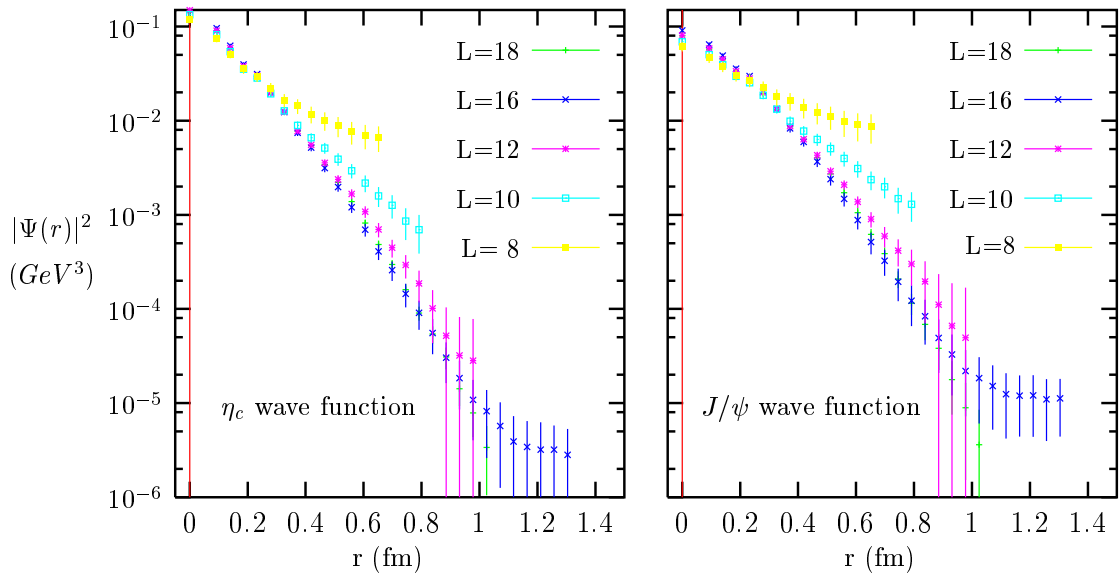}
\caption{Finite volume effects on $\beta=6.0$  pseudoscalar (left) and
vector (right) wave functions.}
\label{fig.fvol}
\end{figure}

A complete finite volume analysis, including the volume dependence of the 
hyperfine splitting in the continuum limit, is beyond the scope of this paper. 
To address  the question of the relevance of finite volume effects, we have 
decided to study instead a set of $L$= 8, 10, 12, 14, 16 and 18 lattices at 
the coarsest, $\beta=6.0$, lattice spacing. Table \ref{t:fvol} presents 
our results for the pseudoscalar and vector masses and the hyperfine splitting
for $\kappa=0.11925$, corresponding, for $L$=18, to a pseudoscalar mass of
$M(\etac)=2949(2)$MeV. Our results are not precise enough to attempt a fit of the 
volume dependence. Still, the variation of the hyperfine splitting amounts 
to at most $4\%$ for $La> 1$ fm. Further indication about the smallness of 
finite volume effects for lattices larger than 1 fm comes from the study of 
the volume dependence of wave functions. Fig.~\ref{fig.fvol} shows the 
results for the set of lattices in Table \ref{t:fvol}. Deviations from the 
large volume behavior are negligible for $La > 1.12$ fm.

\section{Conclusions}

Table \ref{table.all} compiles our result for the $J/\psi-\eta_c$ hyperfine 
splitting together with previous determinations using other lattice formalisms 
\cite{Cppacs2,Trottier}. Our result is obtained with the non-perturbatively
improved clover Dirac operator. The systematic error covers
the difference between 
choosing as reference scale for charm the $\eta_c$ or $J/\psi$ mass
as well as different continuum extrapolations (including or not $\beta=6.0$).
Our final result is $\Delta M(J/\psi-\eta_c) = 77(2)(6)\MeV$.

\begin{table}[htb]
\caption{Hyperfine splitting in MeV from different lattice approaches.
The scale is set by $r_0$ or by the spin-averaged $\overline{P}\!-\! \overline{S}$
splitting. The result quoted as `This work' is obtained from the non-perturbatively
improved clover Dirac operator. }
\vspace*{0.2cm}
\label{table.all}
\newcommand{\m}{\hphantom{$-$}}
\newcommand{\cc}[1]{\multicolumn{1}{c}{#1}}
\renewcommand{\tabcolsep}{0.6pc} 
\renewcommand{\arraystretch}{1.20} 
\begin{tabular}{@{}|l|llll|}
\hline
&This work & CP-PACS \cite{Cppacs2} &
CP-PACS \cite{Cppacs2} &latest NRQCD\cite{Trottier} \\
Formalism &Relativistic   &Relativistic &Relativistic &Non relativistic\\
Lattice &Isotropic & Anisotropic & Anisotropic & \\
Extrapolation &Continuum & Continuum & Continuum & \\
\hline
Scale& $ r_0$ & $r_0$ &
 $\overline{P}- \overline{S}$ & $\overline{P}- \overline{S}$ \\
$\Delta M$& $77(2)(6)$ & $72.6(9)^{(+1.2)}_{(-3.8)}$
&  $85.3(4.4)^{(+5.7)}_{(-2.5)}$& 55(5) \\
\hline
\end{tabular}\\[2pt]
\end{table}

Our value for the quenched hyperfine splitting within the OZI approximation
remains 30$\%$ below the experimental result. Dynamical quark effects are
usually expected to be ${\cal O}(10-20)\%$, which amounts for a large part 
but not all of the discrepancy. Actual dynamical quark effects may turn out to
be larger for this particular quantity but the remaining
discrepancy might also be due to the OZI approximation.

Our study shows the virtue of the ``brute force'' approach:
reliable, consistent continuum extrapolations can be obtained from improved
or non-improved discretizations if the lattice is fine enough. Conversely,
using a coarse lattice implies continuum extrapolations where non-leading
terms may be significant, thus introducing a systematic error which is
very hard to control. From our study, the boundary between these two regimes
sits where one would expect, around $a M_{q \bar{q}} \sim 1$.

Remarkably however, the non-perturbatively improved clover Dirac operator
appears to give reliable extrapolations even starting from $a M_{q \bar{q}} \sim 1.4$.
Non-leading terms remain small, perhaps because leading corrections
${\cal O}(a^2)$ are themselves very small. We consider our
Fig.~\ref{f:extcomp1} as a spectacular advertisement for using this
Dirac discretization.

We believe our result finally closes the long debate on the magnitude of
the quenched charmonium hyperfine splitting (within the OZI approximation). 
While NRQCD or an anisotropic discretization may yield a similar value, 
the reliability of such a result remains questionable: in the first case, 
one deals with an effective theory
where the continuum limit cannot be taken; in the latter, 
the advantage over isotropic lattices in removing ${\cal O} (a m_q)^n$ scaling
violations remains unclear. Until this point is settled, and although
very accurate results can be obtained with anisotropic actions,
this reduction in statistical errors 
can be more than offset by an increase in systematic errors due to possible 
scaling violations. Anisotropic lattices may be very useful in other contexts
of course like for instance at finite temperature. Perhaps our result can be 
used to fine-tune the anisotropic actions involved.

Note finally that the charmonium system is ideally suited for a precision 
study on the current generation of PC clusters. Large, quenched lattices can
be simulated efficiently with little inter-node communication, and high 
accuracy can be obtained at small cost. This computer environment seems well 
suited for a measurement of OZI-effects.


\subsection*{Acknowledgments}

We thank Sara Collins for providing the raw UKQCD data and 
Stefan Sint for discussions and for pointing out the possible 
relevance of OZI suppressed diagrams.
Numerical simulations have been performed on the SX5 at the Research
Center for Nuclear Physics, Osaka University.

\end{document}